\documentstyle[twocolumn,epsf]{article}
\begin{document}
\twocolumn[%
\vspace{1cm}

\centerline{\Large\bf Pressure, Resistance, and Current Activation of}
\centerline{\Large\bf Anisotropic Compressible Hall States}

\ 

\centerline{K. Ishikawa and N. Maeda}

\ 

\centerline{
Department of Physics, Hokkaido University, 
Sapporo 060-0810, Japan}

\ 

\centerline{
\begin{minipage}{14cm}
{\small 
Thermodynamic and electric properties of anisotropic compressible Hall 
states at higher Landau levels are studied using a mean field theory 
on the von Neumann lattice basis. 
It is shown that resistances agree with the recent experiments of 
anisotropic compressible states and the states have negative pressure. 
As implications, the collapse phenomena of the integer quantum Hall 
effect are discussed. }
\end{minipage}
}

\ 

\ 


\ 

]

\vspace{3cm}

In the quantum Hall systems, kinetic energy is quenched and the one particle 
energies are discrete with enormous degeneracy. Its degeneracy per area and 
level spacing are proportional to the magnetic field. 
Electron interaction could change the electron system to have 
completely different properties. 
It is quite interesting if compressible gas states that have continuous 
one particle energies are formed. 
If it happens, pressure and compressibility are expected 
to become negative because the starting kinetic energy which gives positive 
contributions to these thermodynamic quantities in normal electron gas is 
frozen by the magnetic field. 
In this paper we apply Hartree-Fock method and find compressible mean field 
states. 
Their physical properties are shown to agree with recent experiments of 
anisotropic compressible states at higher Landau levels\cite{a,b}. 
Strikingly we find that the gas has in fact a negative pressure and 
compressibility\cite{c}. 
Hence the gas tends to shrink, and strip of compressible states is formed in 
bulk region of finite realistic systems. 
The strip gives unusual electronic resistances to the whole system at finite 
temperature\cite{kawaji,qwe}. 

The injected current flows through the Landau levels below Fermi energy which 
have finite values of wave functions at source and drain regions. 
The strip around Fermi energy has finite width and the wave function 
vanishes at source or drain regions. 
Hence the current does not flow in the strip at zero temperature. 
But at finite temperature the current is activated by the interaction. 
We compute the induced current and resistances at finite temperature. 
We compare them with the recent experiments of breakdown and 
collapse phenomena of the quantum Hall effects with finite injected current. 
Dissipative quantum Hall regime where Hall resistance is quantized in the 
presence of small longitudinal resistance is shown to exist and to play an 
important role in metrology. 

We use von Neumann lattice representation for the Landau levels, which treats 
two dimensional spaces symmetrically and are useful for investigating 
interaction effects and deriving exact relations\cite{e}. 
Formation of compressible states are studied using a mean field theory and 
their physical properties and implications are analyzed. 
Electron operators $a_l({\bf p})$ have momentum, $\bf p$, which is defined 
in the magnetic Brillouin zone $\vert p_i\vert < \pi/a$ $(a=\sqrt{
2\pi\hbar/eB})$ and Landau level index, $l$, 
which determines the energy eigenvalues $E_l=(\hbar eB/m)(l+1/2)$. 
We use the unit of $\hbar=c=a=1$ for simplicity. 
The total Hamiltonian of neutral system $H=H_0+H_1$ is given by 
\begin{eqnarray}
H_0&=&\sum E_l a_l^\dagger({\bf p})a_l({\bf p})\label{ham}\\
H_1&=&\int_{k\neq0}{d^2 k\over(2\pi)^2}
\rho({\bf k}){V({\bf k})\over2}\rho(-{\bf k})
\end{eqnarray}
where $V({\bf k})=2\pi q^2/k$ and the density operator 
$\rho({\bf k})$ is given by
\begin{eqnarray}
\int_{\rm BZ}{d^2p\over(2\pi)^2}\sum_{ll'}a_l^\dagger({\bf p})a_l({\bf p-k})
\langle f_l\vert e^{ik\cdot\xi}\vert f_{l'}\rangle\nonumber\\
\times e^{-ik_x(2p-k)_y/4\pi}.
\end{eqnarray}
Technical details are given in Ref.\cite{e}. 
We obtain anisotropic self-consistent solutions for two point functions and 
the mean field Hamiltonian in the $l$ th Landau level 
defined from Eqs.(1) and (2),
\begin{eqnarray}
\langle a_{l'}^\dagger({\bf p}')a_l({\bf p})\rangle=
\theta(\mu-\epsilon({\bf p}))\delta_{ll'}(2\pi)^2\delta({\bf p}'-{\bf p}),\\
H_{\rm mean}=\int_{\rm BZ}{d^2p\over(2\pi)^2}
\epsilon({\bf p})a_l^\dagger({\bf p})a_l({\bf p})\nonumber\\
+{(a_l,a_l^\dagger)\rm\  independent\ terms}. 
\end{eqnarray}
It would be reasonable to study an anisotropic solution, 
because this matches with the Hall bar geometry. 
The phase factor in Eq.~(3) is a characteristic factor in the strong magnetic 
field and plays important roles. 
In the present translationally invariant mean field solution, 
this phase factor cancels in Eq.~(5). 
The momentum is a good quantum number and the Fermi sea is shown in Fig.(1). 
This solution breaks translational invariance in the momentum, $p_y$, and is 
invariant under translation in the momentum, $p_x$. 
The $p_x$ direction is like integer Hall state which has the energy 
gap of Landau levels but in $p_y$ direction there is no energy gap. 
Density modulation of long wave length along $p_y$ direction stabilizes 
the system, but those along $p_x$ direction does not. 
Due to the phase factor, the density is uniform in y-direction but is 
periodic in x-direction.

\begin{figure}
\centerline{
\epsfysize=1.8in\epsffile{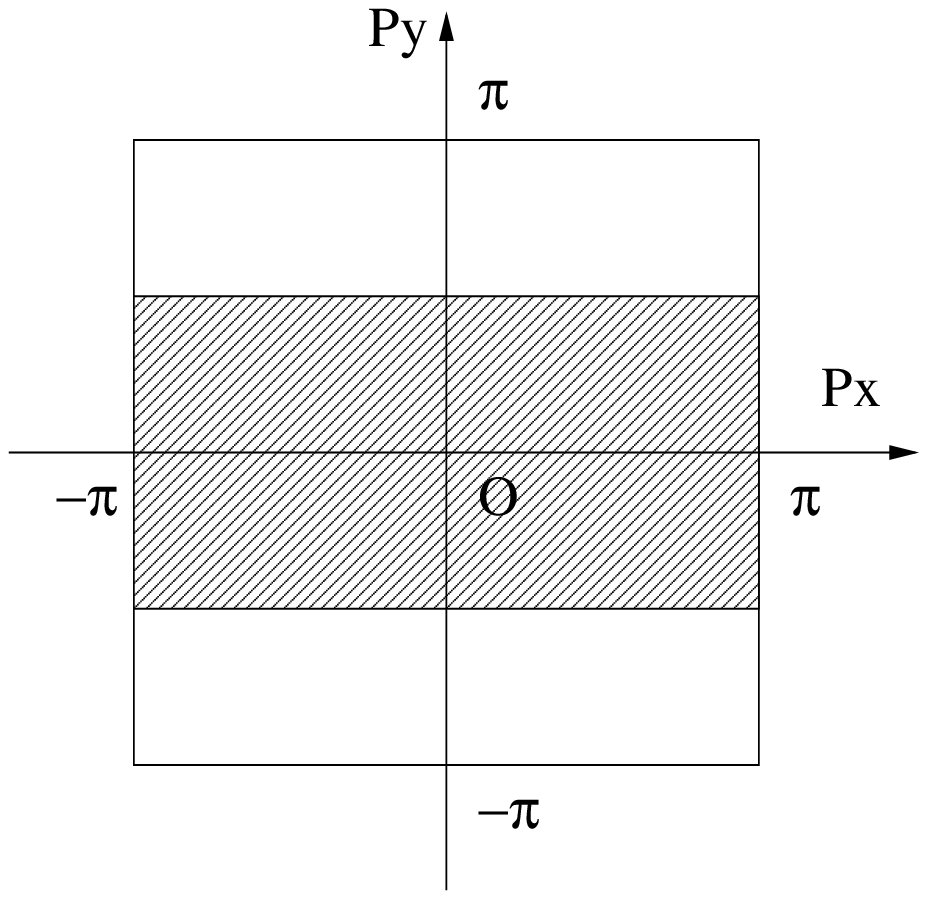}}
Fig~1. The Fermi sea is shown by shaded region in the Brillouin zone. 
\end{figure}

We study infinite systems first. 
The above mean field solution has the energy per particle shown in Fig.(2). 
Pressure and compressibility are computed from these energies and 
given in Figs. (2) and (3). 
We see that both values are negative. 
Since charge neutrality is assumed from the beginning and positive charge 
does not move in the real material, negative pressure does not lead 
to instability. 
We will study the implications of negative pressure later. 
Energy per particle and pressure of an interacting two 
dimensional gas without magnetic field, which has the continuous one particle 
kinetic energy, are shown in the same figures for comparison. 

\begin{figure}[t]
\centerline{
\epsfysize=1.6in\epsffile{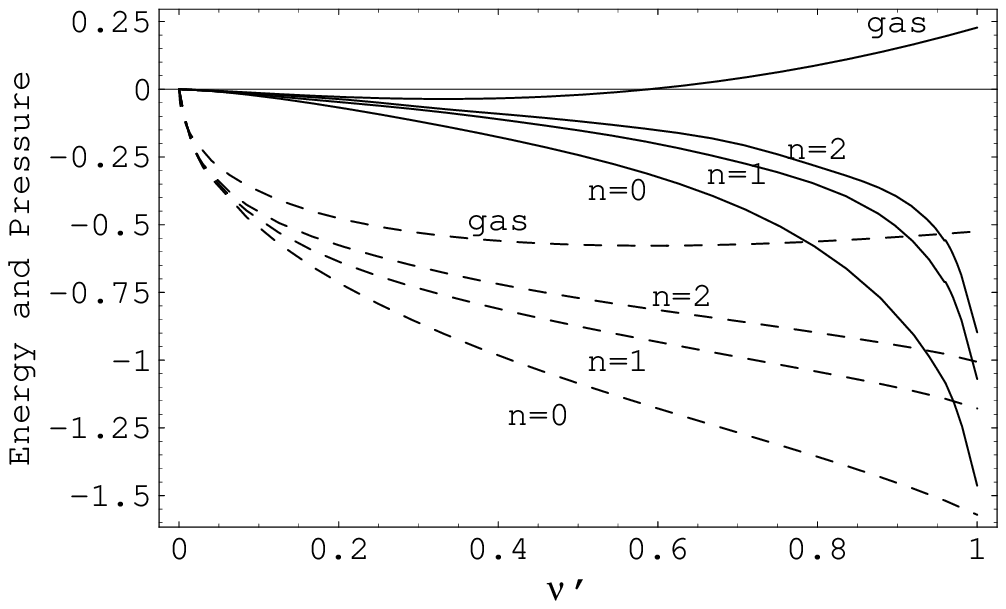}}
Fig~2. Energy per particle (dashed lines) in the unit of $q^2/a$ and 
pressure (solid line) in the unit of $q^2/a^3$ for the filling factor 
$\nu=n+\nu'$. 
Corresponding values for the two-dimensional electron gas without magnetic 
field are also shown (gas) at density$=\nu'/a^2$ for $B=6$T in GaAs. 
\end{figure}

\begin{figure}[t]
\centerline{
\epsfysize=1.6in\epsffile{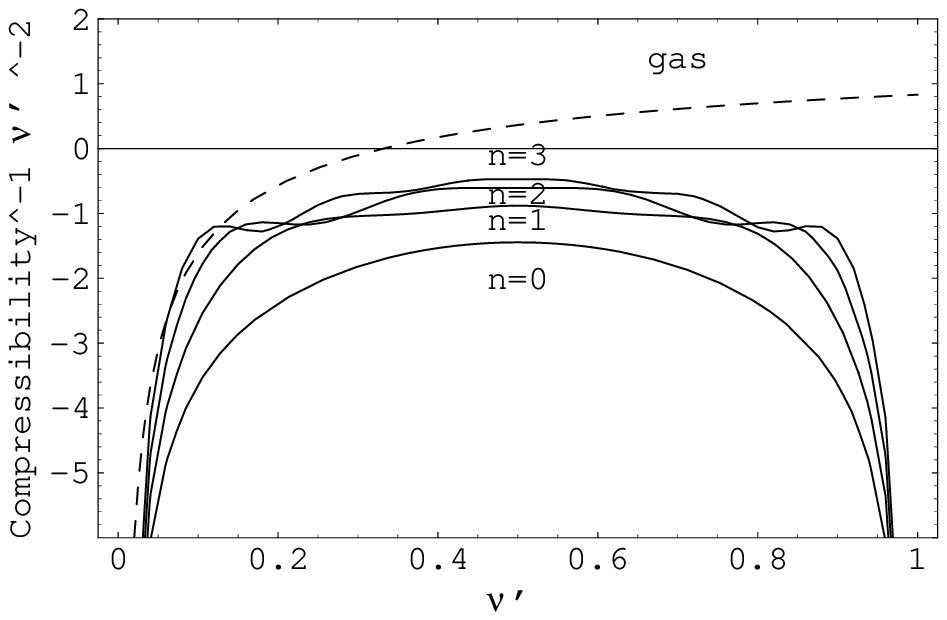}}
Fig~3. Inverse compressiblity times $\nu'^{-2}$ in the unit of 
$q^2/a^3$ for $\nu=n+\nu'$. Dashed line shows the corresponding value of 
the two-dimensional electron gas without magnetic field (same as Fig.~2). 
\end{figure}

We compute resistances of the compressible mean field states. 
From Fig.(1), in $p_x$ direction the empty level is in the next Landau level 
and the system has energy gap. 
Hence longitudinal conductance $\sigma_{xx}$ vanishes. 
The electron system is like one dimensional system and $\sigma_{yy}$ is 
given by that of Landauer formula, $e^2/h$. 
The Hall conductance is given by the topological formula\cite{e},
\begin{equation}
\sigma_{xy}={e^2\over h}{1\over24\pi^2}\int{\rm tr}({\tilde S}(p)
d{\tilde S}^{-1}(p))^3. 
\end{equation}
where ${\tilde S}(p)$ is the propagator in the current basis in which 
the standard Ward-Takahashi identity is satisfied. 
The propagator for the compressible states are written by using the momentum 
dependent energy $\epsilon({\bf p})$ as, 
$
S^{-1}(p)_{ll'}=\{p_0-(E_l+\epsilon({\bf p}))\}\delta_{ll'}. 
$
The Hall conductance becomes, 
\begin{equation} 
\sigma_{xy}={e^2\over h}(n+\nu').
\end{equation}
The values of $\sigma_{ij}$ agree with the recent experiments 
of anisotropic compressible states at $\nu=n+1/2$\cite{a,b}.

In the real experiments, are used the semiconductors of finite sizes. 
They have impurities and confining potentials. 
Owing to random impurities, one particle states are localized and have finite 
spatial extensions except those at the center of the Landau level. 
They contribute to transport if their localization lengths are 
same order as or larger than the system's sizes. 
There are three system's sizes, length of the system $L_s$, 
width at Hall probe region $L_{w1}$, 
and width at potential probe region $L_{w2}$. 
Corresponding to three system's sizes, three energies 
in which localization lengths agree with the system's sizes are defined. 
In the center of Landau levels, wave functions are extended. 
We consider the energy regions between the center of the $l$-th Landau level 
$E_l$ and the boundary of the l-th Landau level with the $l+1$-th Landau
level. 
Let the energies, $E_s$, $E_{w1}$, and $E_{w2}$ be the energy values 
where the localization lengths, $\lambda(E)$, agree with the three system's 
sizes, $\lambda(E_s)=L_s$, $\lambda(E_{wl})=L_{w1}$, $\lambda(E_{w2})=L_{w2}$. 
States in the energy range $E<E_s$ are extended states, states in the energy 
range $E_s<E<E_{w1}$ bridge from one edge to the other edge at the potential 
probe region and at the Hall probe region, states in the energy range 
$E_{w1}<E<E_{w2}$ bridge from one edge to the other edge at the potential 
probe region. 
Finally states in the energy range $E_{w2}<E$ are localized. 
We call these energy regions the extended state region, collapse region, 
pre-collapse region or dissipative quantum Hall regime, and localized state 
region or quantum Hall regime, respectively. 
Reasons why we use particular names for the last two regions will become 
clear later. 
The system's lengths for a typical Hall bar and energy regions 
are shown in Fig. (4). 

\begin{figure}
\centerline{
\epsfysize=1.2in\epsffile{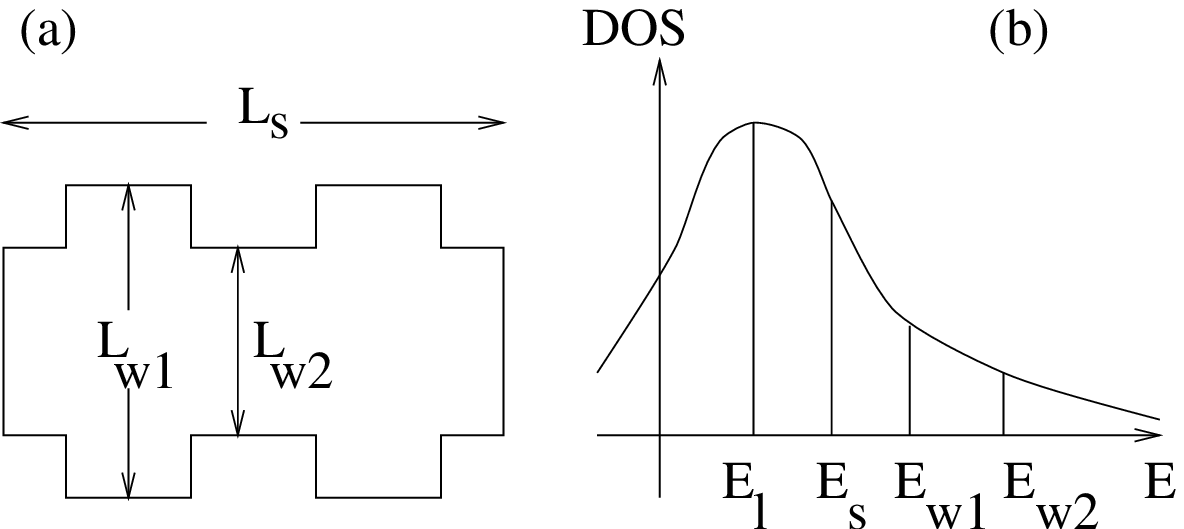}}
Fig~4. (a) Schematical view of a Hall bar. 
(b) Density of state (DOS) for the $l$ th Landau level. 
\end{figure}
Compressible states from degenerate Landau levels are extended and cover whole 
spatial area from one end to another end and have negative pressure and 
compressibility. 
Positive charges are composed of ions and do not move in the materials. 
Hence once the negative charges shrink, Coulomb energy makes the energy of 
the system higher. 
At one point, pressure effect and the Coulomb force from positive charge ions 
are balanced. 
The shape is determined by minimizing the total energy. 
If the filling of the compressible gas is around half filling, 
the effects of Coulomb interaction are negligible. 
But if the filling is slightly away from integer, the negative pressure 
effects are strong and strip is formed. 
The width of strip is determined from the balance between negative pressure 
and Coulomb interaction of electrons with neutralizing background charge. 

Wave functions in collapse regime and dissipative quantum Hall regime bridge 
from one edge to another edge and are of stripe shape. 
Coulomb interaction due to positive charge ions makes them shrink further. 
Hence wave functions are modified slightly. 
Thus the strip is formed in the bulk regions. 

The strip is located in the bulk region and its wave functions vanish at 
source and drain regions. The current does not flow through compressible 
gas strip at zero temperature even though states have energies of Fermi 
energy. 
At finite temperature, the current is activated into the strip from the 
current carrying extended states at lower Landau levels. 
The electric current in the strip is induced from the higher order effect 
shown in Fig.(5), and is given by  
\begin{equation}
J=q^2\beta (eE_H/\gamma) e^{-\beta\Delta E} F_l(\beta eE_H/\gamma),
\label{j}
\end{equation}
where $1/\beta$ is temperature, $\Delta E$ is the energy gap, 
$E_H$ is Hall electric field, and $1/\gamma$ is an enhancement factor 
due to localized states. 
$F_l$ is a function which does not vary so largely and is calculated 
numerically. 
The energy gap $\Delta E$ in Eq.~(\ref{j}) is given by 
$\Delta E=E_F-E_s-eE_H/2\gamma$ in the systems of finite inject current
\cite{f}. 

\begin{figure}
\centerline{
\epsfysize=0.8in\epsffile{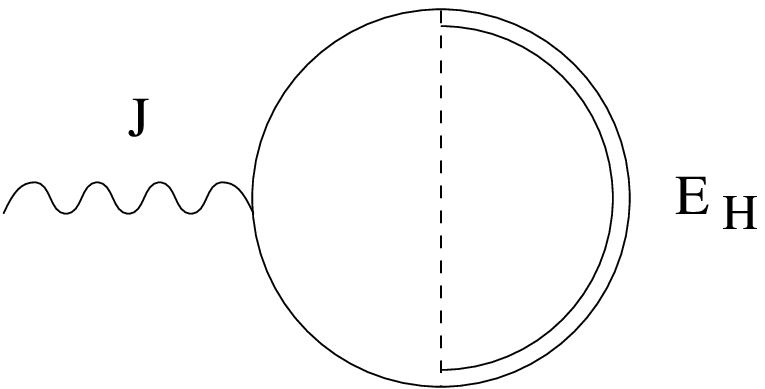}}
Fig~5. Feynmann diagram for the induced current. 
The dashed line stands for an interaction effect and double line 
stands for the propagator in the presence of the Hall electric field. 
\end{figure}

The longitudinal resistance and Hall resistance are computed and are 
summarized in Table.(1). 
The Fermi energy varies with magnetic field, electron density, or injected 
current. 
By tuning them, we are able to adjust the Fermi energy in dissipative 
quantum Hall regime or quantum Hall regime. 
The Hall resistance is quantized exactly as inverse of 
integer multiple of $e^2/h$ in both regions. 
In the former regime the longitudinal resistance does not vanish but behaves 
like activation type with finite energy gap. 
The heating from dissipation determines the temperature at strip regions 
($1/\beta_s$) and the energy gap 
depends on Hall electric field in lower Landau levels. 
In the latter regime, the longitudinal resistance vanishes at zero 
temperature and becomes activation type at finite temperature. 
The energy gap is determined from the difference between extended state's 
energy and the Fermi energy. 

\vspace{0.5cm}
\noindent
Table 1 Resistances are summarized in each regions. 
$\Delta\rho_{xy}$ is a deviation from the quantized value and $1/\beta_s$ 
is a temperature at strip regions. 

\ 

\noindent
\begin{tabular}{cccc}
\hline\hline 
&collapse&pre-collapse& quantum Hall\\
\hline 
$\Delta\rho_{xy}$ & $\propto e^{-\beta_s\Delta E}$& 0 & 0 \\
$\rho_{xx}$ &$\propto e^{-\beta_s\Delta E}$&$\propto e^{-\beta_s\Delta E}$
& 0 \\\hline\hline
\end{tabular}
\vspace{0.5cm}

In summary, it is shown that 
anisotropic states have unusual properties and 
the dissipative quantum Hall regime exists at finite temperature. 
Collapse phenomena observed by Kawaji et al. are shown to be 
understandable theoretically. 

This work was partially supported by the special Grant-in-Aid for 
Promotion of Education and Science in Hokkaido University provided by the 
Ministry of Education, Science, Sport, and Culture, the Grant-in-Aid for 
Scientific Research on Priority area (Physics of CP violation) 
(Grant No. 12014201).

\end{document}